\documentclass[12pt]{article}

\pdfoutput=1
\usepackage{graphicx}
\usepackage{amssymb}
\usepackage{amsmath}
\usepackage{color}
\usepackage{cite}
\usepackage{subfigure}
\usepackage{bm}
\usepackage[colorlinks=true,linkcolor=blue,citecolor=blue]{hyperref}

\begin{document}


\begin{titlepage}

\begin{flushright}
\end{flushright}

\vskip 2cm

\begin{center}

{\Large \bf
Resonant production of dark photons from axion without a large coupling
}

\vspace{1cm}

Naoya Kitajima$^{\,a,b}$
and
Fuminobu Takahashi$^{\,b}$ \\

\vskip 1.0cm

{\it
$^a$Frontier Research Institute for Interdisciplinary Sciences, Tohoku University, Sendai, Miyagi 980-8578, Japan\\[2mm]
$^b$Department of Physics, Tohoku University, 
Sendai, Miyagi 980-8578, Japan 
}

\vskip 1.0cm

\begin{abstract}
Dark photons could be produced resonantly by the oscillating axion field in the early universe. This resonant production mechanism has been used in various contexts, including dark photon dark matter and primordial magnetic field production. However, for this resonant production to work in an expanding universe, a large axion-dark photon coupling is required, which is not easy to realize in terms of model building and requires the introduction of many charged fermions and/or the complex clockwork mechanism. In this paper, we present a new scenario that efficiently produces dark photons from the axion with a much smaller coupling. This is possible by modifying the dynamics of axion and significantly delaying the onset of oscillations, as in the so-called trapped misalignment mechanism. As a specific example, we consider models in which dark photon production occurs efficiently despite the small axion-dark photon coupling by temporally trapping an axion in a wrong minimum and releasing it after the Hubble parameter becomes much smaller than the axion mass. In this scenario, it is expected that the polarization asymmetry of dark photons and gravitational waves generated from dark photons will be significantly reduced.
\end{abstract}

\end{center}

\end{titlepage}

\newpage

\vspace{1cm}

\section{Introduction} \label{sec:intro}
Dark photons are known as a plausible candidate for physics beyond the Standard Model~\cite{Holdom:1985ag}. Recently they have attracted much attention as a candidate for dark matter, which can be probed  by various experiments/observations via the kinetic mixing with the standard model photons.
For reviews, see e.g. Refs.~\cite{Fabbrichesi:2020wbt,Caputo:2021eaa}.

The production mechanism of dark photon dark matter has been studied in the literature, such as the production from the inflationary fluctuations \cite{Graham:2015rva,Sato:2022jya}, the resonant  production from the axion oscillation \cite{Agrawal:2018vin,Co:2018lka,Bastero-Gil:2018uel},
gravitational production during reheating \cite{Ema:2018ucl}, misalignment production of coherent oscillation \cite{Nakayama:2019rhg,Nakayama:2020rka,Kitajima:2023fun}, production from cosmic strings \cite{Long:2019lwl,Kitajima:2022lre}, and resonant production from dark Higgs \cite{Dror:2018pdh}.
In this paper, we focus on the resonant production of dark photons from the axion condensate.

Dark photons can be resonantly produced from the coherent oscillation of axion field~\cite{Garretson:1992vt}, and if the axion-dark photon coupling is large enough, the dark photon production occurs efficiently until the production  is saturated. Then the axion dynamics is significantly affected by the produced dark photons through the back reaction. In particular, when applied to the QCD axion, the axion abundance can be reduced to be several \% as a consequence of the dark photon production \cite{Kitajima:2017peg} (see also Ref.~\cite{Agrawal:2017eqm}). Stochastic gravitational waves are also sourced by the produced dark photons, which predict circular polarization asymmetry of the gravitational wave background as a unique signal of this scenario \cite{Adshead:2018doq,Machado:2018nqk,Namba:2020kij,Salehian:2020dsf,Kitajima:2020rpm,Ratzinger:2020oct,Co:2021rhi,Madge:2021abk}. In addition, if the dark photon mass is generated by the Higgs mechanism, 
the dark photons produced by the process can effectively stabilize the dark Higgs field at its origin until late times, where the hidden U(1) symmetry is restored. Then, the dark Higgs can lead to an additional short-term inflation phase~\cite{Kitajima:2021bjq} or the early dark energy~\cite{Nakagawa:2022knn}.  Focusing on the coupling with the standard model photons instead of dark photons, the primordial large-scale magnetic field may also be produced by the same mechanism (e.g. \cite{Fujita:2015iga}).\footnote{When electrons or other charged particles are present, they can hamper the growth of the gauge fields. This makes it harder to generate magnetic fields using the resonant production method.}

While the resonant production of dark photons is a very interesting phenomenon and is being investigated in a variety of contexts, it
necessitates a large axion-dark photon coupling, which, in turn, requires a non-trivial UV completion~\cite{Agrawal:2018vin}. The situation is exacerbated when the dark gauge coupling  is small, because the anomalous axion-dark photon coupling is considered to be proportional to the gauge coupling squared. For example, one way to realize such a large axion-dark photon coupling is to introduce many symmetry-breaking scalar fields as in the clockwork mechanism~\cite{Higaki:2016yqk} (see also Refs.~\cite{Kim:2004rp,Choi:2014rja,Higaki:2014qua,Higaki:2015jag,Choi:2015fiu,Kaplan:2015fuy,Giudice:2016yja,Farina:2016tgd}). Alternatively, one may introduce a large number of charged fermions that run in the loop diagram.

In this paper, we show a novel possibility to efficiently produce dark photons from coherent oscillations of the axion for a much smaller coupling. This is due to a modification of the axion dynamics. Specifically, we consider an additional potential which traps the axion in early times. Such a potential can significantly delay the commencement of axion oscillations  compared to the conventional scenario where the axion starts to oscillate when the Hubble parameter becomes comparable to the axion mass. 
The trapping effect on the axion has been investigated in the context of the QCD axion with an extra Peccei-Quinn symmetry breaking term~\cite{Kawasaki:2015lpf,Nomura:2015xil,Takahashi:2015waa,Higaki:2016yqk,Nakagawa:2020zjr,Nakagawa:2021nme,Jeong:2022kdr}, a lighter QCD axion~\cite{DiLuzio:2021gos}, and the axion or axion-like particle undergoing a first-order phase transition~\cite{Nakagawa:2022wwm}. It was shown that the trapped misalignment mechanism~\cite{Higaki:2016yqk,Nakagawa:2020zjr,Jeong:2022kdr} can significantly increase or decrease the axion abundance, depending on the minimum in which the axion is trapped, as compared to the conventional misalignment mechanism~\cite{Preskill:1982cy,Abbott:1982af,Dine:1982ah}. Here we show that, by temporally trapping the axion in a wrong minimum and delaying the onset of oscillations as in the trapped misalignment mechanism,
dark photons are resonantly produced  even for the axion-dark photon couplings that are comparable to or smaller than ${\cal O}(1)$.

\section{Model} \label{sec:model}

\begin{figure}[tp]
\centering
\subfigure[Step-function trapping]{
\includegraphics [width = 7.5cm, clip]{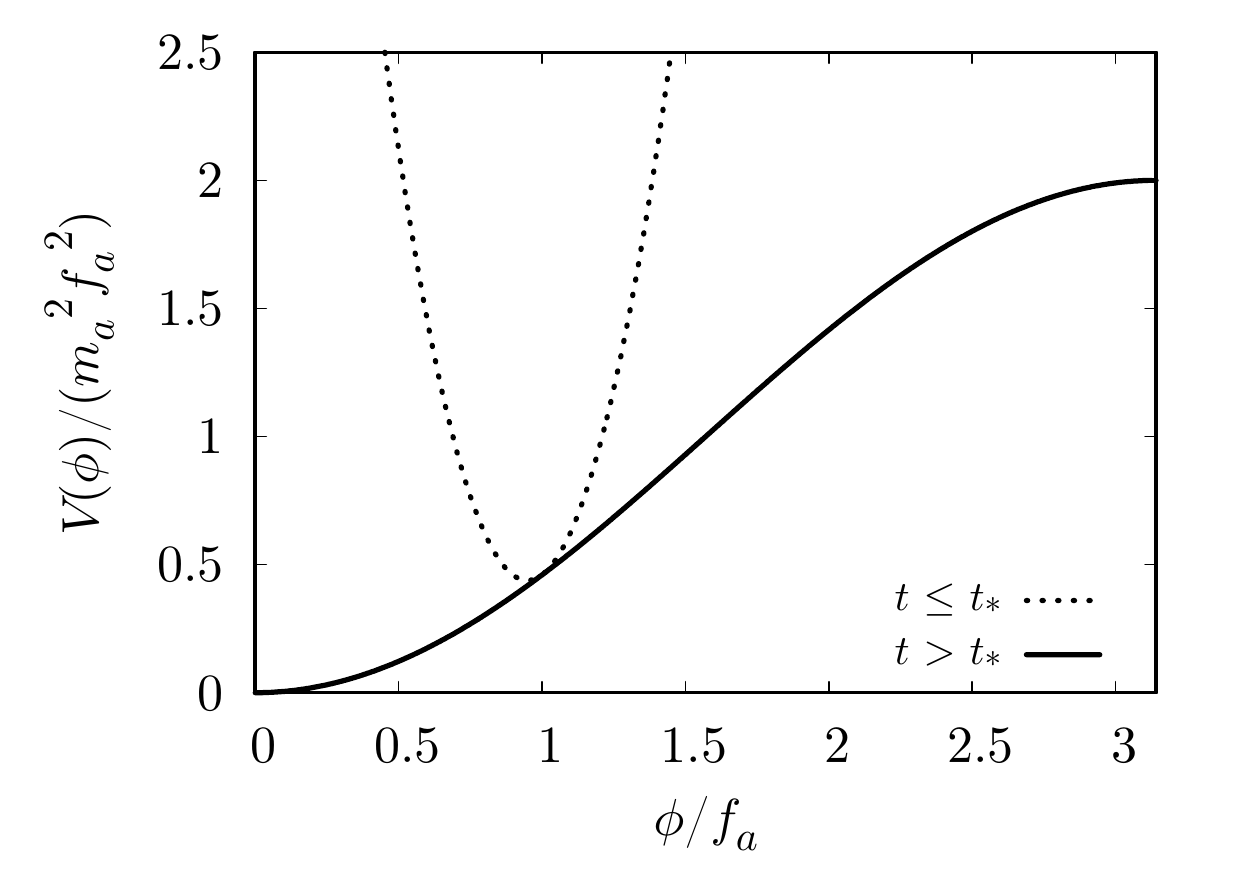}
\label{subfig:potential_a}
}
\subfigure[Trapped QCD axion]{
\includegraphics [width = 7.5cm, clip]{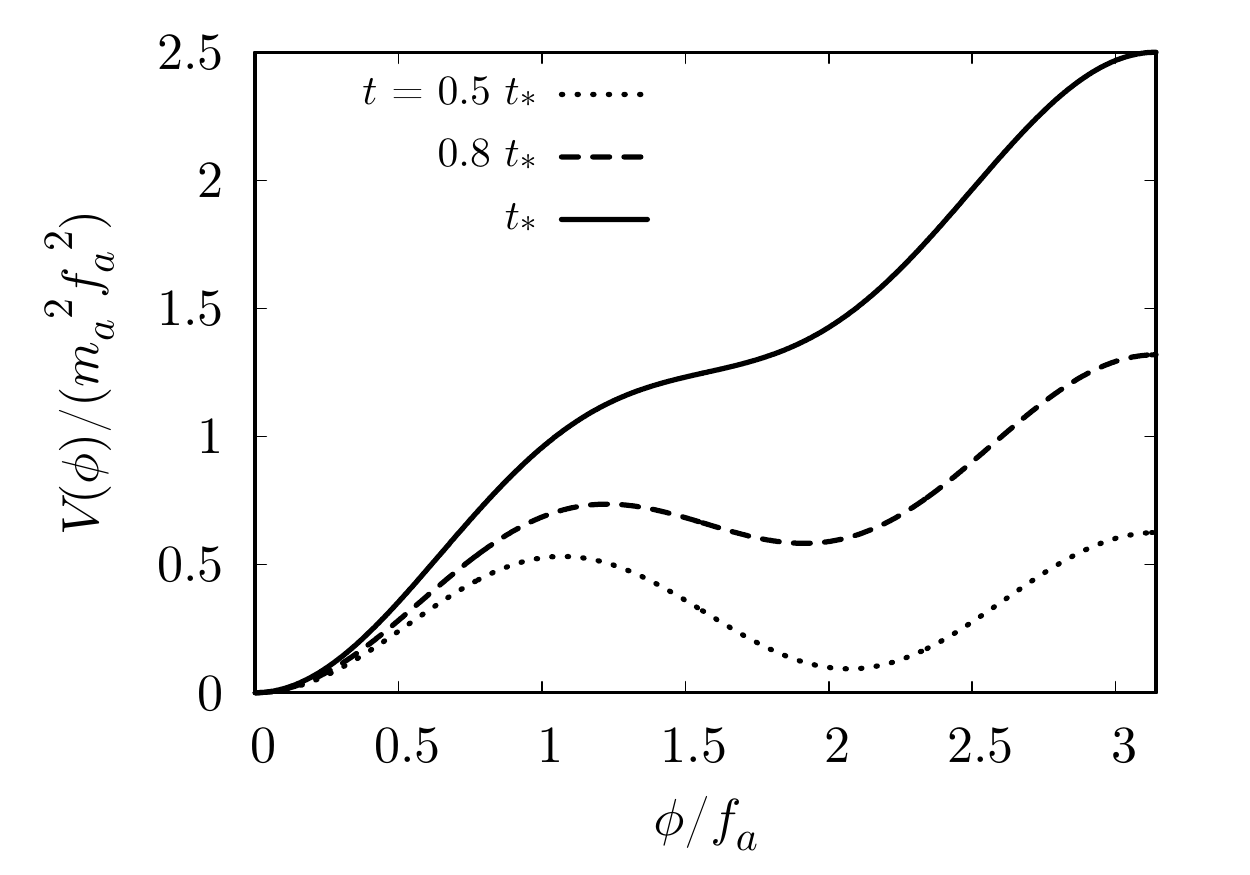}
\label{subfig:potential_b}
}
\caption{
Schematic illustration of the two potentials, a trapping by a step function of time with $m_* = 4m_a$ and $\phi_* = f_a$ (left) and the trapped QCD axion with $N_H=3$ and $\Lambda_H^2=0.5m_{a0}f_a$ (right).
}
\label{fig:potential}
\end{figure}

Let us consider the system of axion ($\phi$) and dark photon ($A_\mu$). The Lagrangian is given by
\begin{align}
{\cal L} = \frac{1}{2}\partial_\mu\phi \partial^\mu \phi - V(\phi) -\frac{1}{4} F_{\mu\nu} F^{\mu\nu}+\frac{1}{2}m_A^2 A_\mu A^\mu-\frac{\beta}{4f_a} \phi F_{\mu\nu} \tilde{F}^{\mu\nu},
\end{align}
where $F_{\mu\nu} = \partial_\mu A_\nu - \partial_\nu A_\mu$ is the field strength tensor, $\tilde{F}^{\mu\nu} = \epsilon^{\mu\nu\rho\sigma} F_{\rho\sigma}/2\sqrt{-g}$ with $\epsilon^{0123}=1$ is its dual, $m_A$ is the mass of the dark photon, $f_a$ is the axion decay constant and $\beta$ is the axion-dark photon coupling.
Here we consider the following axion potential,
\begin{align}
V(\phi) = m_a^2 f_a^2 \left[ 1-\cos\bigg(\frac{\phi}{f_a} \bigg) \right]+V_{\rm trap}(\phi),
\end{align}
with $V_{\rm trap}$ being the potential which traps the axion in early times.
For instance, one can consider the following trapping potential as a toy model,
\begin{align}
V_{\rm trap} (\phi) = \frac{1}{2}m_*^2(\phi-\phi_*)^2 \theta(t_*-t),
\end{align}
where $\theta(t)$ is the step function with respect to the cosmic time $t$,  $t_*$ is the time when the trapping ends, and $m_*~(>m_a)$ is the mass stabilizing the axion around $\phi=\phi_*$. We also consider the model motivated by the trapped QCD axion model \cite{Jeong:2022kdr} with the potential given by
\begin{align}
V(\phi) = m_a(t)^2 f_a^2 \left[ 1-\cos\bigg(\frac{\phi}{f_a}\bigg)\right]+V_{\rm trap}(\phi),
\end{align}
where the time-dependent mass is given by
\begin{align}
m_a(t) = \begin{cases} m_{a0}\bigg(\cfrac{t}{t_*}\bigg)^{b/2}  &~~\text{for}~~t<t_* \\[3mm] m_{a0} &~~\text{otherwise}  \end{cases}.
\end{align}
The trapping potential is given by
\begin{align} \label{eq:VtrapQCD}
V_{\rm trap}(\phi) = \Lambda_H^4 \left[1-\cos\bigg(\frac{N_H \phi}{f_a}\bigg)\right],
\end{align}
where $N_H$ is an integer.
In case of the QCD axion, the time dependent mass is expressed as $m_{a0}(\Lambda/T)^b$ with $\Lambda \sim O(0.1)\,{\rm GeV}$ being the QCD scale and $b\simeq 4$.
In this model, first, the axion falls into one of the minima of the trapping potential (\ref{eq:VtrapQCD}).
Fig.~\ref{fig:potential} shows the schematic illustration of the potential with the above two cases.

From the Lagrangian, one obtains the following equations of motion,
\begin{gather} \label{axion_eom}
\ddot{\phi} + 3H\dot{\phi} -\frac{\nabla^2 \phi}{a^2} + V'(\phi) = -\frac{\beta}{4f_a} F_{\mu\nu} \tilde{F}^{\mu\nu}, \\[2mm]
\ddot{\bm{A}} + H\dot{\bm{A}}-\frac{\nabla^2 \bm{A}}{a^2} - \frac{\beta}{f_a a}\left(\dot\phi \nabla \times \bm{A} -\nabla\phi \times(\dot{\bm{A}}-\nabla A_0) \right)= 0,
\end{gather}
where $a$ is the scale factor, $H=\dot{a}/a$ is the Hubble parameter, with the overdot denoting the derivative with respect to the cosmic time.
One can decompose the gauge field by circular polarization states,
\begin{align}
\bm{A}(t,\bm{x}) = \sum_{\lambda = \pm} \int\frac{d^3k}{(2\pi)^3} \bm{e}_\lambda(\bm{k}) e^{i\bm{k}\cdot\bm{x}} A_\lambda(t,\bm{k}),
\end{align}
where $\bm{e}_\pm(\bm{k})$ are circular polarization vectors satisfying $\bm{k} \cdot \bm{e}_{\pm} (\bm{k}) = 0$ and $\bm{k} \times \bm{e}_{\pm}(\bm{k}) = \mp i |\bm{k}| \bm{e}_{\pm}(\bm{k})$.
Then, assuming the homogeneous background including the axion field, the evolution equation for each mode is simplified as
\begin{align} \label{eq:gauge_field_eom}
\ddot{A}_{\pm} + H\dot{A}_{\pm} +\left(m_A^2 + \frac{k^2}{a^2} \mp  \frac{k}{a}\frac{\beta \dot{\phi_0}}{f_a} \right) A_{\pm} = 0,
\end{align}
where $\phi_0$ is the background value of the axion coherent oscillation.
Note that it shows that the dark photon experiences  tachyonic instabilities depending on the coupling $\beta$ and the sign of $\dot{\phi}_0$.
Soon after the beginning of the axion oscillation, the dark photon field can be produced exponentially.
The equation of motion for the axion coherent oscillation is 
\begin{align}
\ddot{\phi}_0 + 3H\dot{\phi}_0 + V'(\phi_0) = \frac{\beta}{f_a} \langle \bm{E} \cdot \bm{B} \rangle,
\label{axionEOMbg}
\end{align}
where the right-hand-side represents the backreaction by the produced dark photons and the angular bracket represents an average over volume.
It can be calculated in terms of the Fourier modes of the circular polarization states as,
\begin{align}
\langle \bm{E} \cdot \bm{B} \rangle = -\frac{1}{a^3} \int \frac{d^3k}{(2\pi)^3} \frac{k}{2} \frac{d}{dt} \big( |A_+|^2-|A_-|^2 \big).
\end{align} 
The electric and magnetic fields can be represented respectively by $E_i = -\dot{A_i}/a$, $B_i = \epsilon_{ijk} \partial_j A_k /a^2$ under the temporal gauge ($A_0=0$).
The energy density of the dark photon can be calculated as follows
\begin{align}
\rho_A^{(\pm)} = \frac{1}{2} \int \frac{dk}{k} \left[ {\cal P}^{(\pm)}_E(k) + {\cal P}^{(\pm)}_B(k) \right],
\end{align}
with the power spectra given by
\begin{align}
{\cal P}^{(\pm)}_E(k) = \frac{k^3 |\dot{A}_{\pm}|^2}{2 \pi^2 a^2},~~ {\cal P}^{(\pm)}_B(k) = \frac{k^5 |A_{\pm}|^2}{2 \pi^2 a^4}.
\end{align}

In the case with small $\beta$, and a sufficient long trapping of the axion before the onset of oscillations, the gauge field is amplified by the narrow resonance rather than the tachyonic instability. When it begins to oscillate,  the homogeneous axion field evolves as $\phi(t) = \theta_i f_a (a_{\rm osc}/a)^{3/2} \cos(m_a t)$. Then Eq.~(\ref{eq:gauge_field_eom}) can be expressed as the following Mathieu equation,
\begin{align}
    \frac{\partial^2\mathcal{A}_{\pm}}{\partial z^2}+\left[p-2q\cos(2z)\right]\mathcal{A}_{\pm} = 0,
\end{align}
where we have defined $z\equiv m_a t/2+\pi/4$, $\mathcal{A}_{\pm} \equiv a^{1/2}A_{\pm}$, $p=4k^2/(am_a)^2+4m_A^2/(m_a)^2$ and $q = \mp 2k\beta \theta_i(a_{\rm osc}/a)^{3/2} /(am_a)$, and we have assumed $H \ll m_a$. 
Here, we are interested in the case of the narrow resonance, $|q| < 1$, with $\beta\theta_i \lesssim 1$.  Then, the parametric amplification occurs in the first resonance band satisfying $1-|q| < p < 1+|q|$ \cite{Kofman:1994rk,Kofman:1997yn}. Neglecting the dark photon mass for simplicity, the resonance condition reads
\begin{align}
    1-\frac{\beta\theta_i}{2}\bigg(\frac{a_{\rm osc}m_a}{2k}\bigg)^{3/2} \lesssim \frac{am_a}{2k} \lesssim 1+\frac{\beta\theta_i}{2}\bigg(\frac{a_{\rm osc}m_a}{2k}\bigg)^{3/2},
\end{align}
where we dropped higher order terms of $q$ since $|q| \ll 1$.
Note that the center of the instability band is at $k/a=m_a/2$ and the band width is proportional to $\beta\theta_i$.
Therefore the time interval during the resonant amplification can be estimated as
\begin{align}
    z_{\rm res} = \frac{1}{2} m_a(t_{\rm end}-t_{\rm osc}) \approx \frac{\beta\theta_i m_a}{H_{\rm osc}}\left(\frac{2k}{a_{\rm osc}m_a} \right)^{-3/2},
\end{align}
where $t_{\rm end}$ is the time at the end of the resonance.
Neglecting the backreaction to the axion, the gauge field is maximally amplified as $\mathcal{A}_{\pm} \propto \exp(\mu z_{\rm res})$ with $\mu \simeq |q|/2$ being the growth rate. Thus, the total growth exponent can be computed as follows
\begin{align} \label{eq:muz}
    \mu z_{\rm res} \simeq \frac{1}{4}|q| m_a(t_{\rm end}-t_{\rm osc}) \simeq \frac{1}{2\sqrt{2}}\left(\frac{k}{a_{\rm osc} m_a}\right)^{-1/2} \frac{(\beta\theta_i)^2m_a}{H_{\rm osc}}.
\end{align}
Note that the wave number giving the maximum enhancement is $k \approx a_{\rm osc}m_a/2$ 
since the momentum of produced dark photons should kinematically satisfy $k/a_{\rm osc} \gtrsim m_a/2$.

In the conventional scenario with $H_{\rm osc} \sim m_a$, the growth exponent is $(\beta\theta_i)^2$. Then the significant dark photon production never occurs for a small coupling ($\beta \ll 1$). In the trapped axion scenario, on the other hand, the growth exponent can be enhanced by a factor of $m_a/H_{\rm osc}$. Thus, the explosive dark photon production is possible even for a small coupling if the onset of the axion oscillation is delayed significantly.

In the next section, we will numerically solve the equations of motion for the axion and dark photon fields and examine whether the trapping effect of the axion can induce resonant production of dark photons even for small $\beta$. Although we could perform lattice simulations in principle, our goal only requires us to estimate the growth rate of dark photon fields. Therefore, we will restrict ourselves to the linear analysis and ignore the feedback of the produced dark photons on the axion dynamics.

\section{Numerical results}
We solve the equation of motion (\ref{eq:gauge_field_eom}) for the dark photon numerically, assuming a homogeneous background that includes the axion field. Since our interest is in the growth of dark photon fields for various axion potentials, we neglect the back reaction and set the r.h.s. of Eq.~(\ref{axionEOMbg}) to zero. We consider three scenarios of axion dynamics: one without the trapping potential $V_{\rm trap}$, one with a step function of time, and one with a trapped QCD axion model.

\begin{figure}[tp]
\centering
\subfigure[$\beta=10$ without trap]{
\includegraphics [width = 7.5cm, clip]{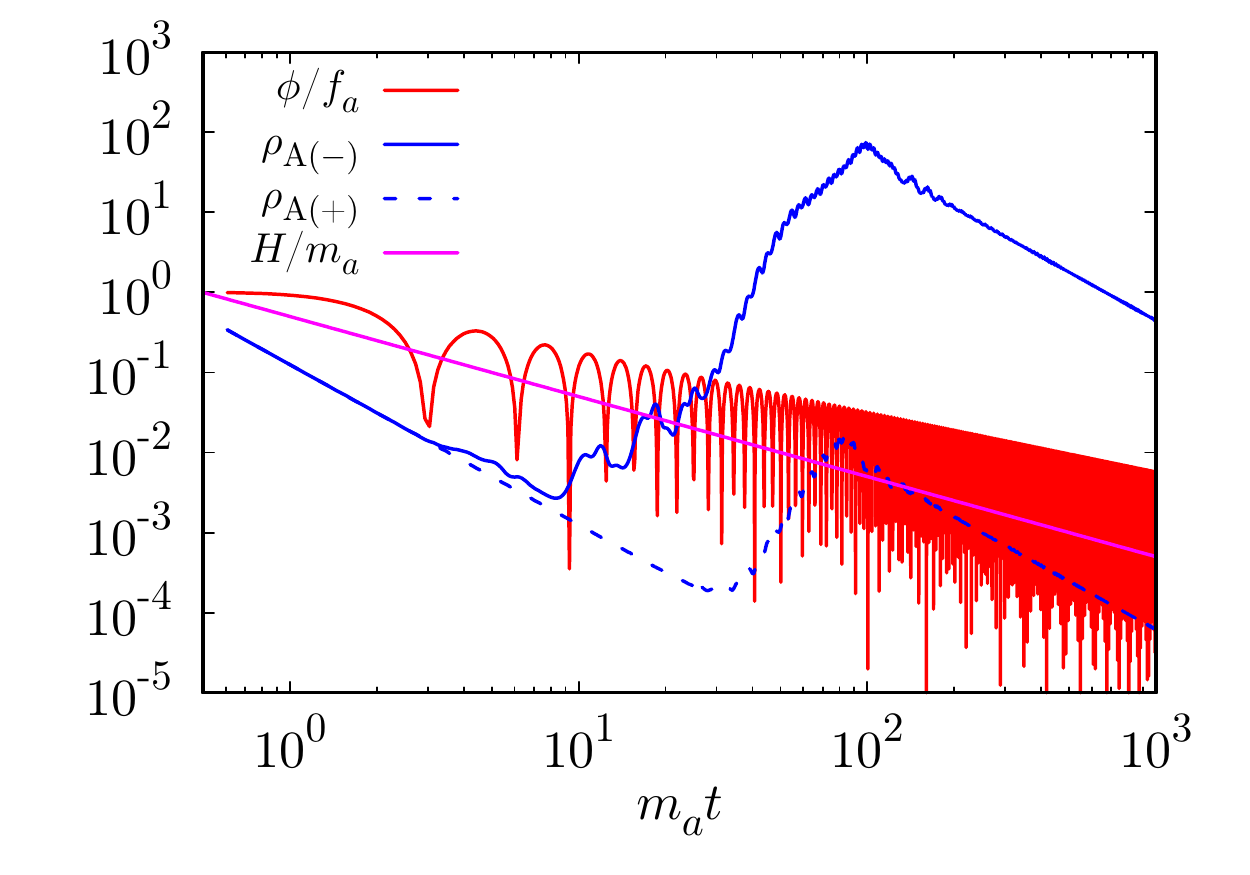}
\label{subfig:evolve_a}
}
\subfigure[Step-function trapping, $\beta=1$, $m_at_*=50$]{
\includegraphics [width = 7.5cm, clip]{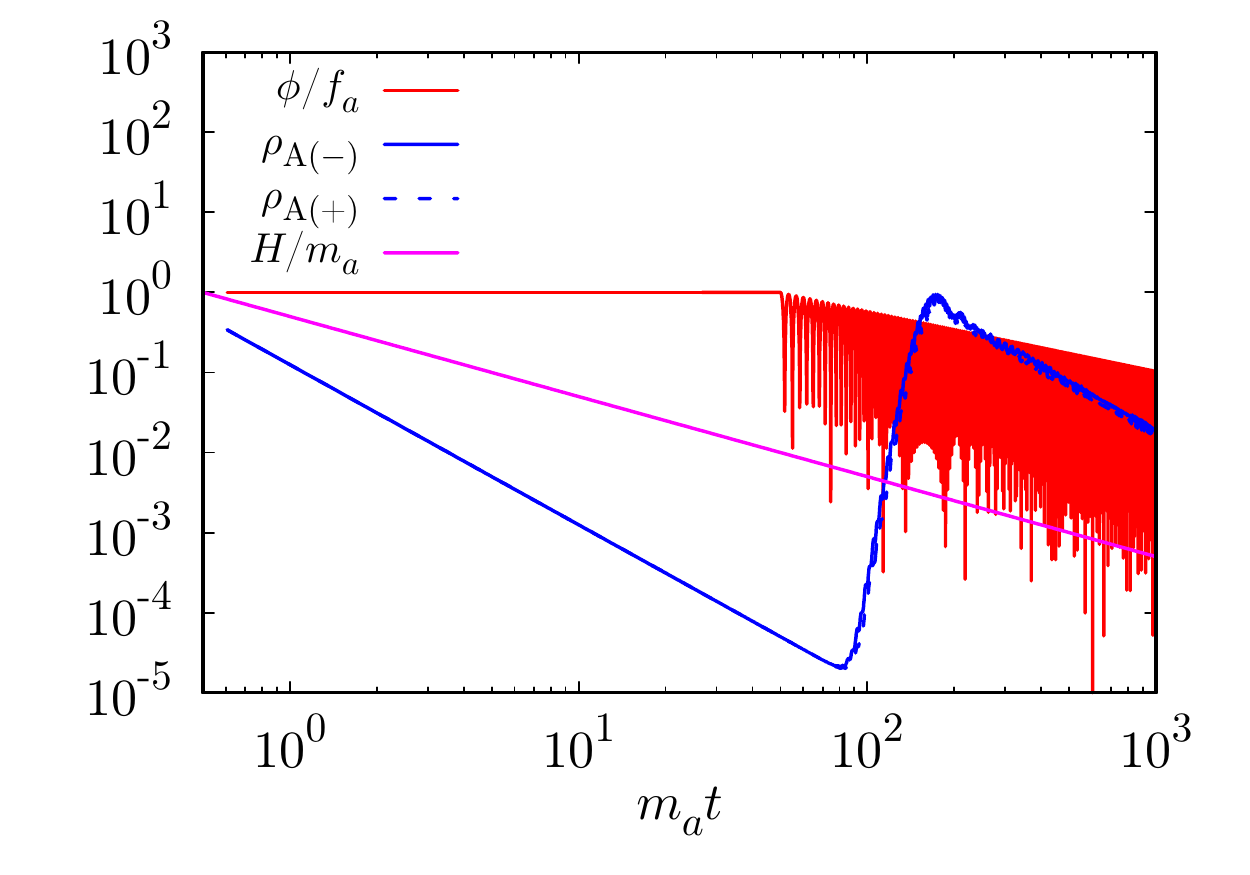}
\label{subfig:evolve_b}
}
\subfigure[Trapped QCD axion, $\beta=1$, $m_at_*=50$]{
\includegraphics [width = 7.5cm, clip]{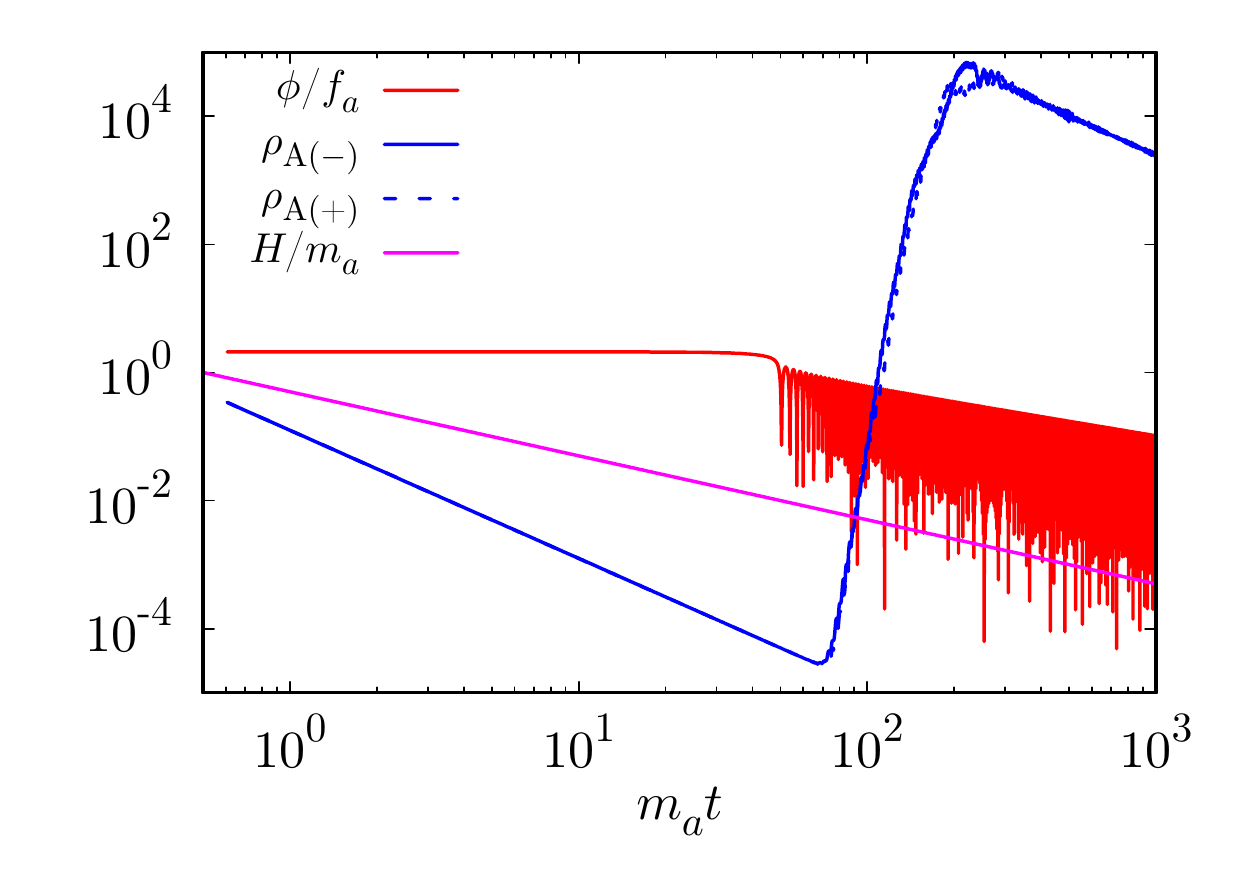}
\label{subfig:evolve_c}
}
\caption{
Time evolution of the energy density of the dark photon (blue), the axion field value (red) and the Hubble parameter (maganta). Solid and dashed blue lines corresponds respectively to the minus and plus circular polarization modes.
We have taken $m_A=0$ and $\beta=10$ (top-left), $\beta=1$ and $m_at_*=50$ (top-right and bottom), $m_*=4m_a$ (top-right), $N_H=3$ and $\Lambda_H^2 = 0.5m_a f_a$ (bottom).
}
\label{fig:evolve}
\end{figure}

In Fig.~\ref{fig:evolve}, we show the time evolution of the axion field value and the energy density of each circular polarization mode of the dark photon field for the three cases mentioned above. In the case without trapping shown in panel (a), it can be seen that a large asymmetry in the circular polarization is generated in the dark photons produced in the first few oscillations, due to the cosmic expansion. Although the back reaction to the axion is not included,  the production of dark photons stops at some point because instability band becomes smaller due to cosmic expansion. On the other hand, in the two trapped cases shown in panels (b) and (c), the axion oscillation period is consistently much shorter than Hubble time from right after the trapping ends, and therefore both plus and minus modes are equally generated, resulting in no circular polarization asymmetry. This is in contrast to the conventional case (a) where there is a large difference between two modes of generated dark photons. In both cases (b) and (c), we set $\beta=1$ , but it can be seen that resonant production of dark photons occurs efficiently.

\begin{figure}[tp]
\centering
\subfigure[without trap]{
\includegraphics [width = 7.5cm, clip]{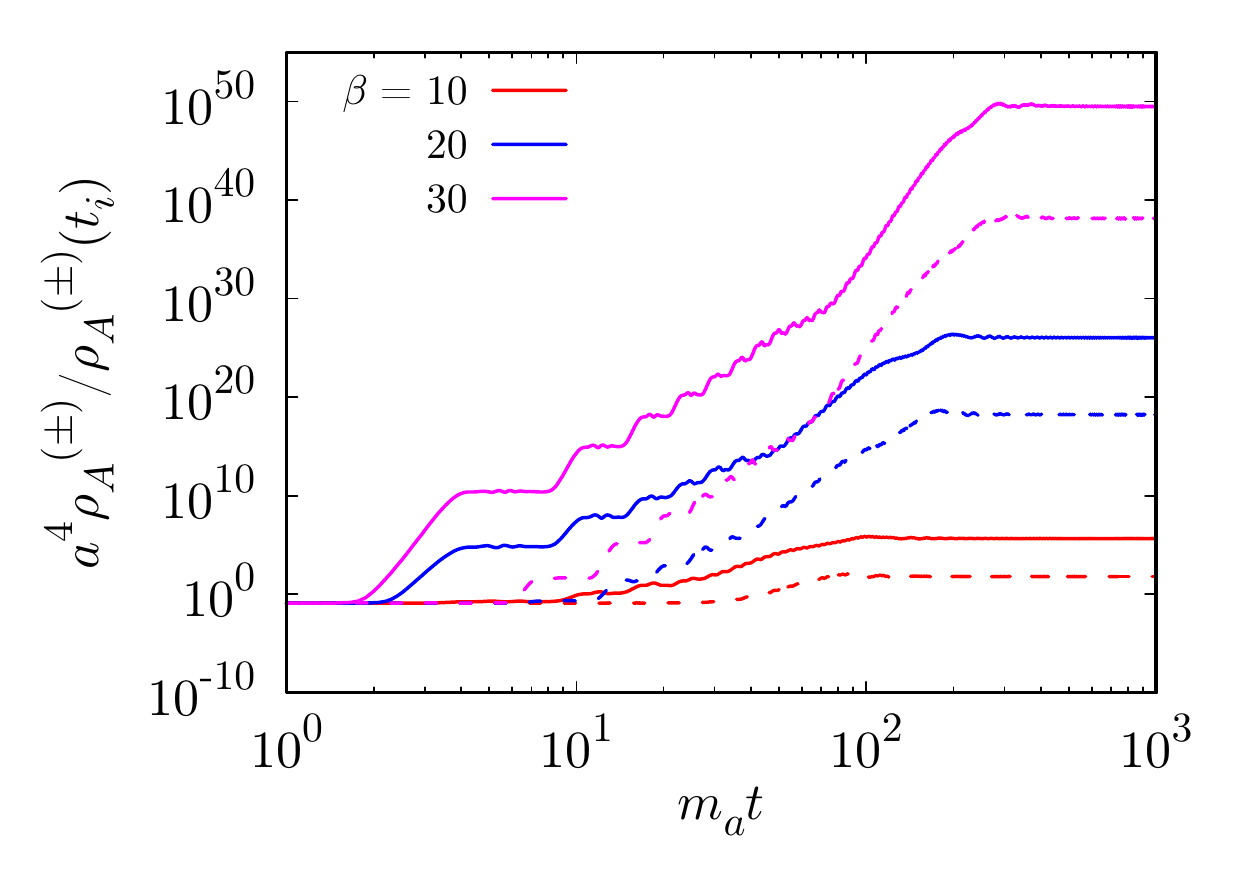}
\label{subfig:ev_growth_notrap}
}
\subfigure[Step-function trapping, $\beta=1$]{
\includegraphics [width = 7.5cm, clip]{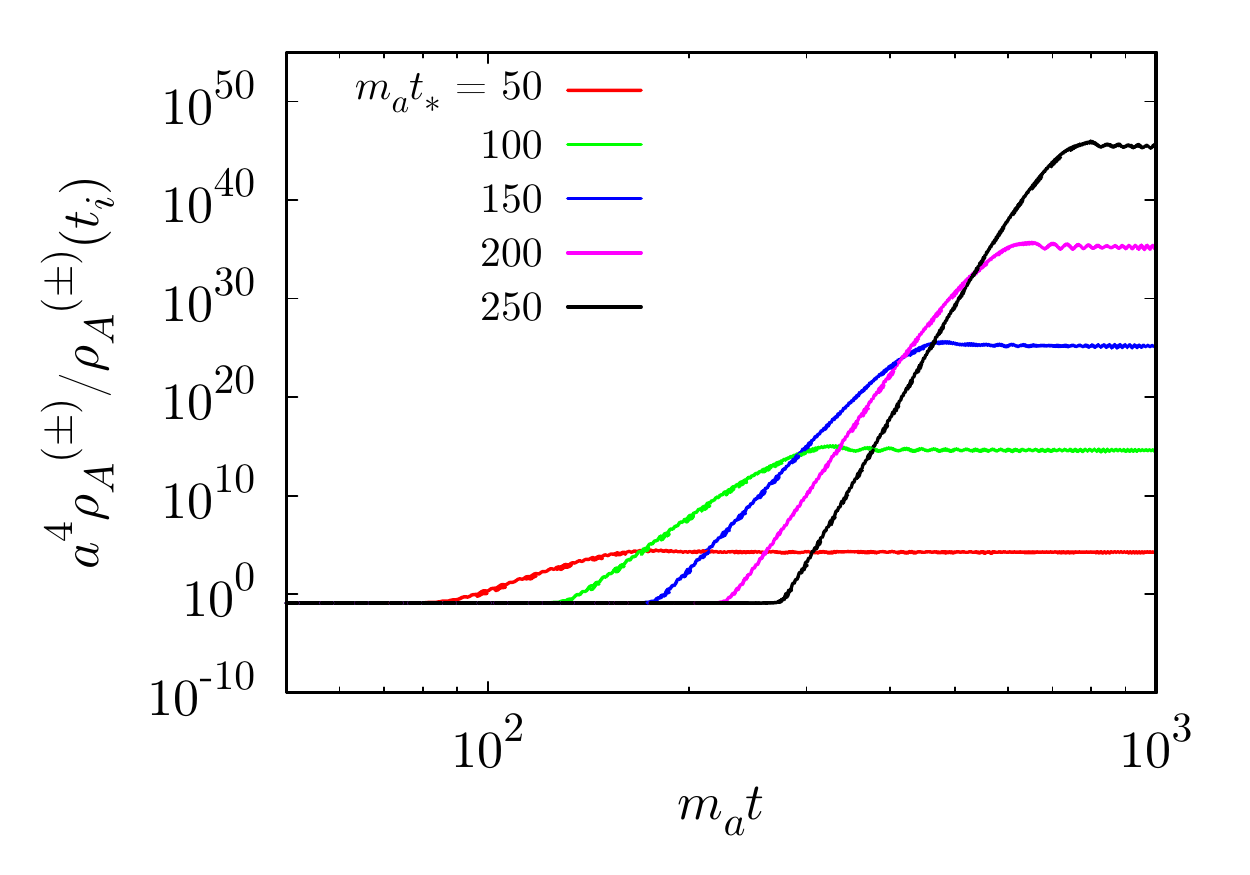}
\label{subfig:ev_growth_a}
}
\subfigure[Trapped QCD axion, $\beta=1$]{
\includegraphics [width = 7.5cm, clip]{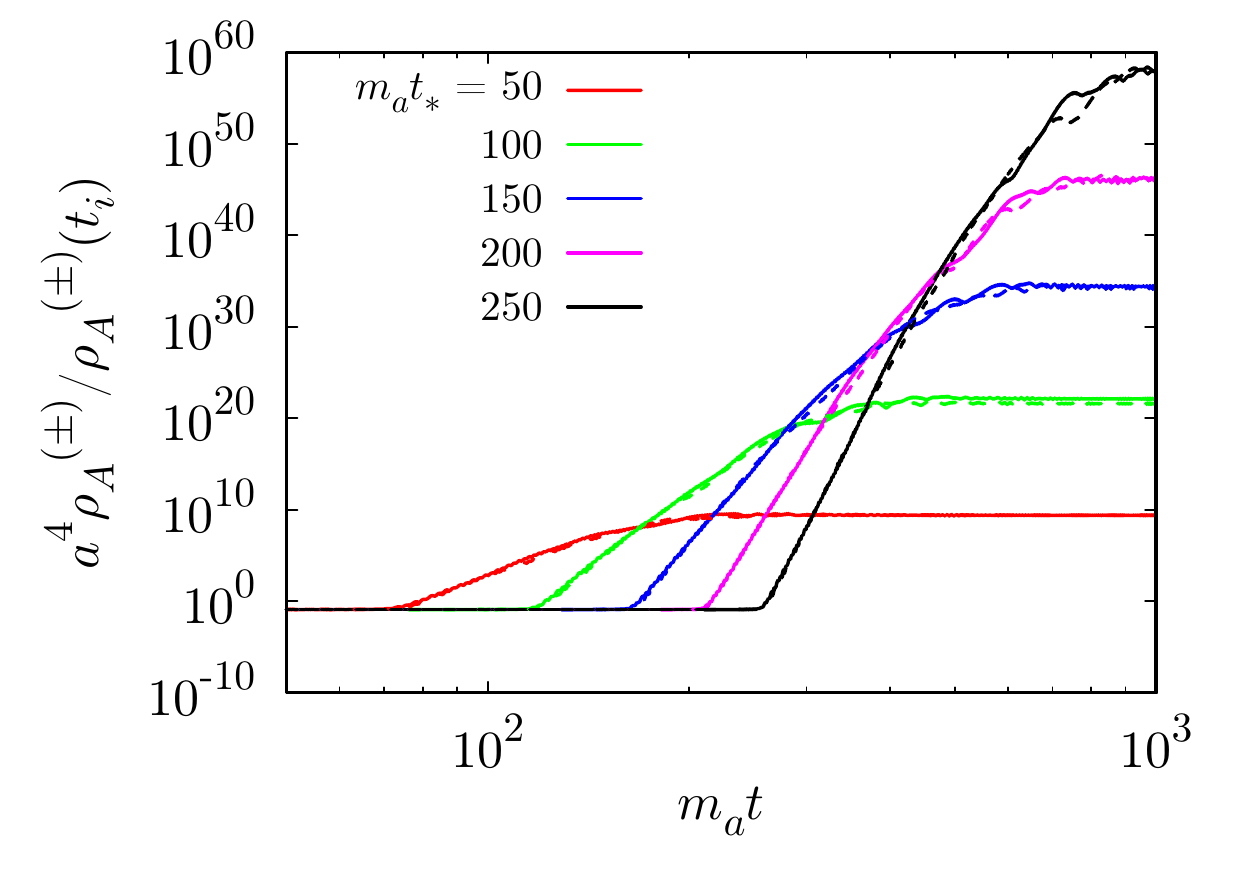}
\label{subfig:ev_growth_b}
}
\caption{
Time evolution of the comoving energy density of the dark photon normalized by the initial value. The thick and dashed line corresponds respectively to the minus and the plus circular polarization modes.
We set $m_A=0$, $m_*=4m_a$ (top-right), $N_H=3$ and $\Lambda_H^2 = 0.5m_a f_a$ (bottom) and neglected the backreaction to the axion.
}
\label{fig:ev_growth}
\end{figure}

In Fig.~\ref{fig:ev_growth}, we show  how much the energy density of dark photons increased compared to the initial value in the three cases, by changing $\beta$ and the end time of trapping $t_*$. In the case without trapping, it can be seen that dark photons are generated more efficiently as $\beta$ increases. On the other hand, in two trapped cases with $\beta=1$, it can be seen that dark photons are generated more efficiently as the end time of trapping becomes later. As before the circular polarization asymmetry is suppressed in the trapped cases.

\begin{figure}[tp]
\centering
\subfigure[$\beta = 30$ without trap]{
\includegraphics [width = 7.5cm, clip]{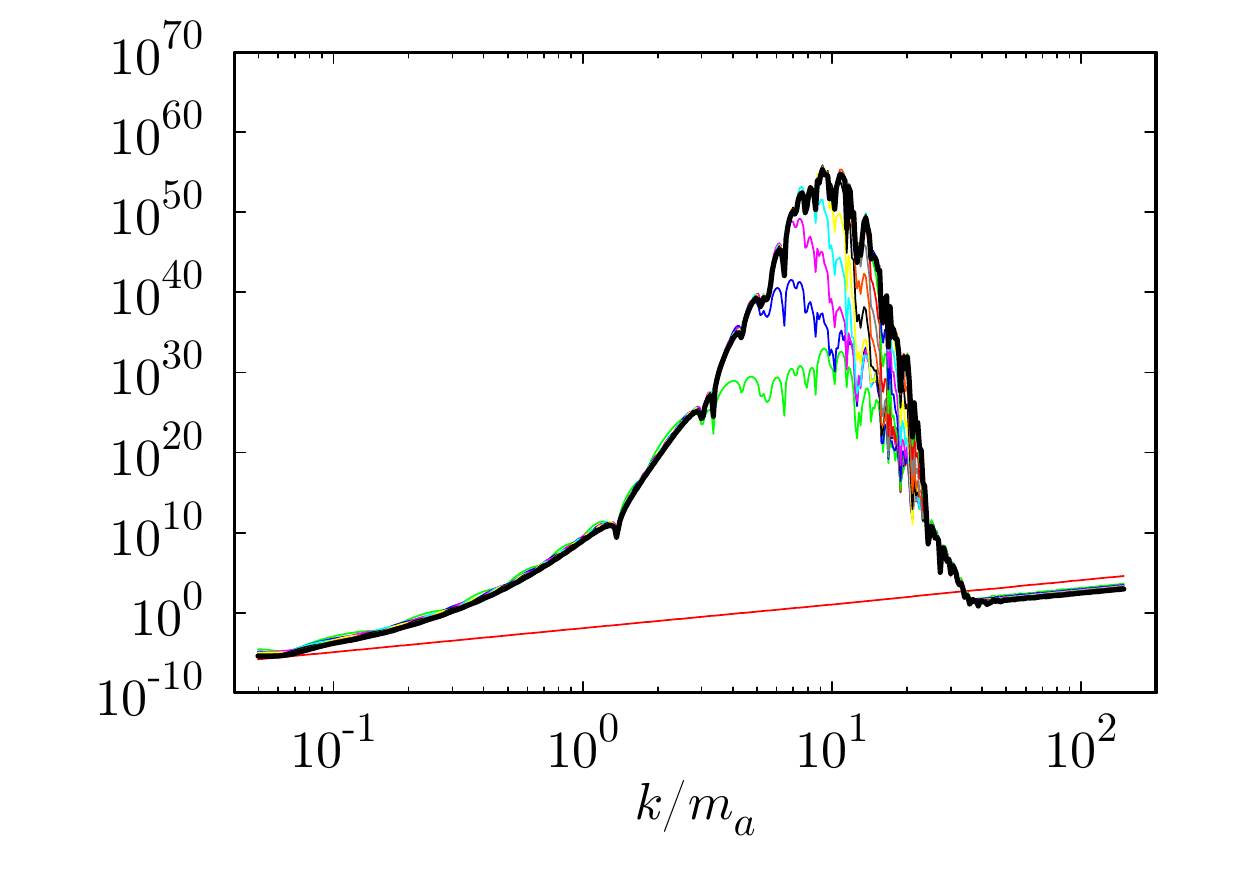}
\label{subfig:spectrum_a}
}
\subfigure[Step-function trapping, $\beta=1$, $m_at_*=250$]{
\includegraphics [width = 7.5cm, clip]{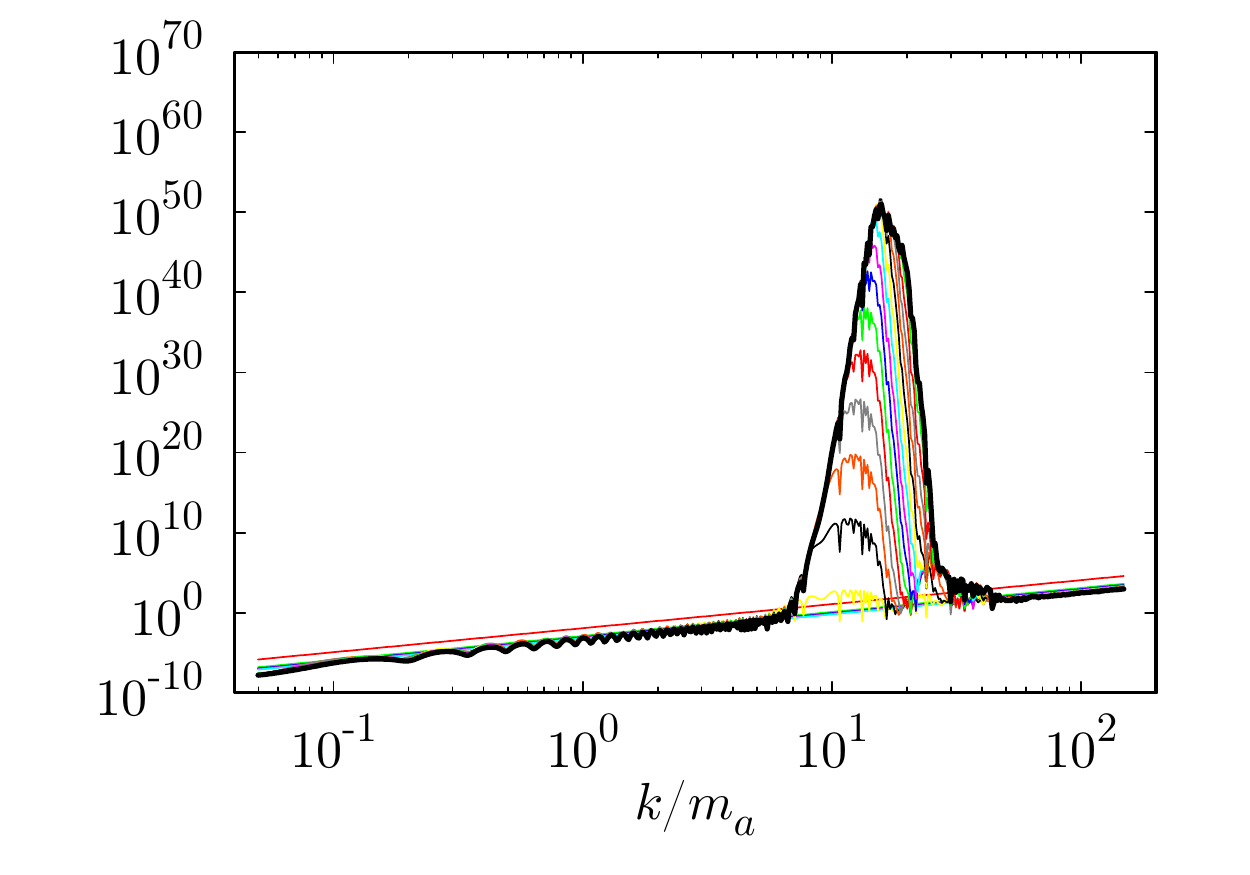}
\label{subfig:spectrum_b}
}
\subfigure[Trapped QCD axion, $\beta=1$, $m_at_*=250$]{
\includegraphics [width = 7.5cm, clip]{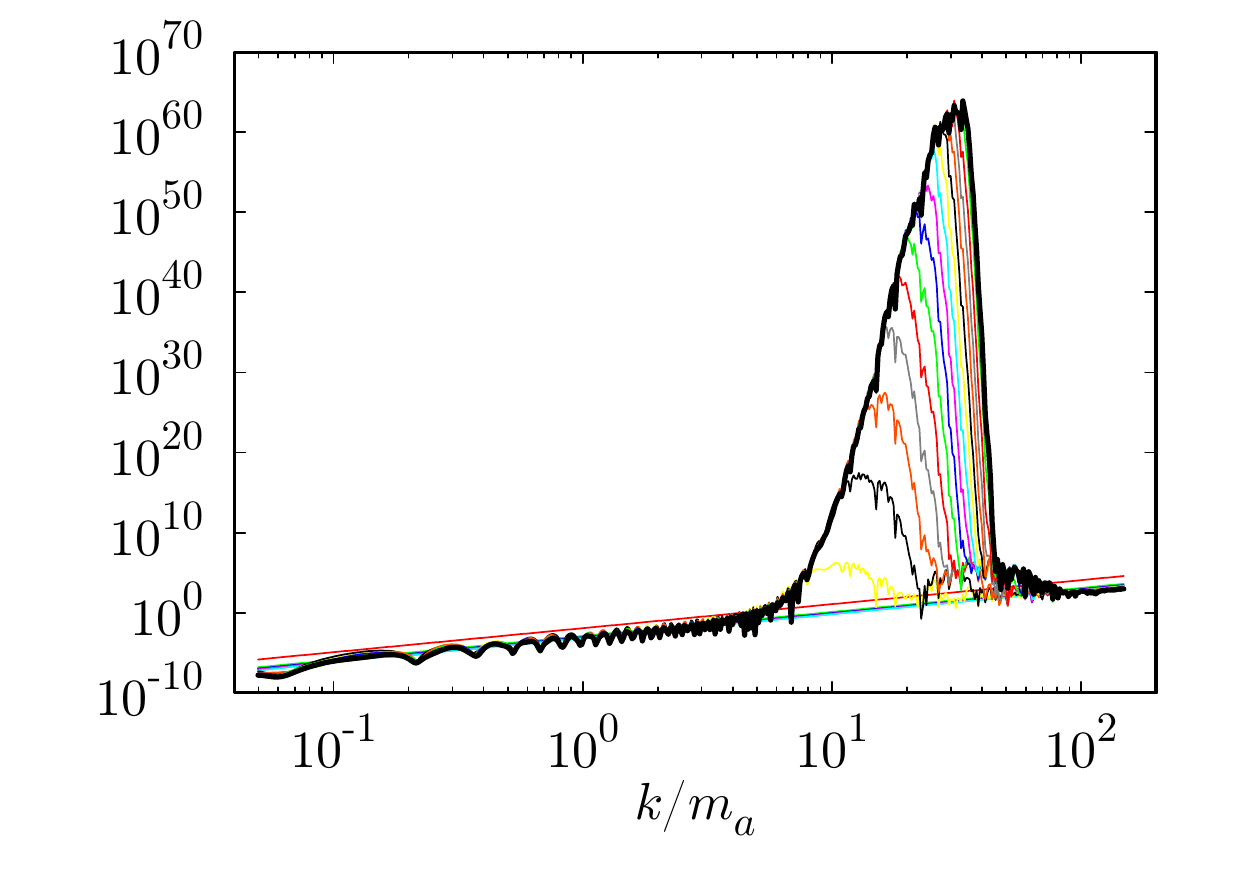}
\label{subfig:spectrum_c}
}
\caption{
Time evolution of the spectrum of comoving number density of the dark photon. 
We set $m_A=0$, $m_*=4m_a$ (top-right), $N_H=3$ and $\Lambda_H^2 = 0.5m_a f_a$ (bottom) and neglected the backreaction to the axion.
The time evolves from bottom to top and the thick black line corresponds to the final time $m_a t = 1000$.
}
\label{fig:spectrum}
\end{figure}

We also show the time evolution of the spectrum of dark photons for each case in Fig.~\ref{fig:spectrum}. Compared to when there is no trap, we can see that the spectrum becomes sharper. This is what is expected from the previous discussion that resonance is narrow. Since the horizontal axis is the comoving wavenumber,  we can also see that the fastest growing mode moves to a larger comoving wavenumber as times goes.

\begin{figure}[tp]
\centering
\subfigure[Without trap]{
\includegraphics [width = 7.5cm, clip]{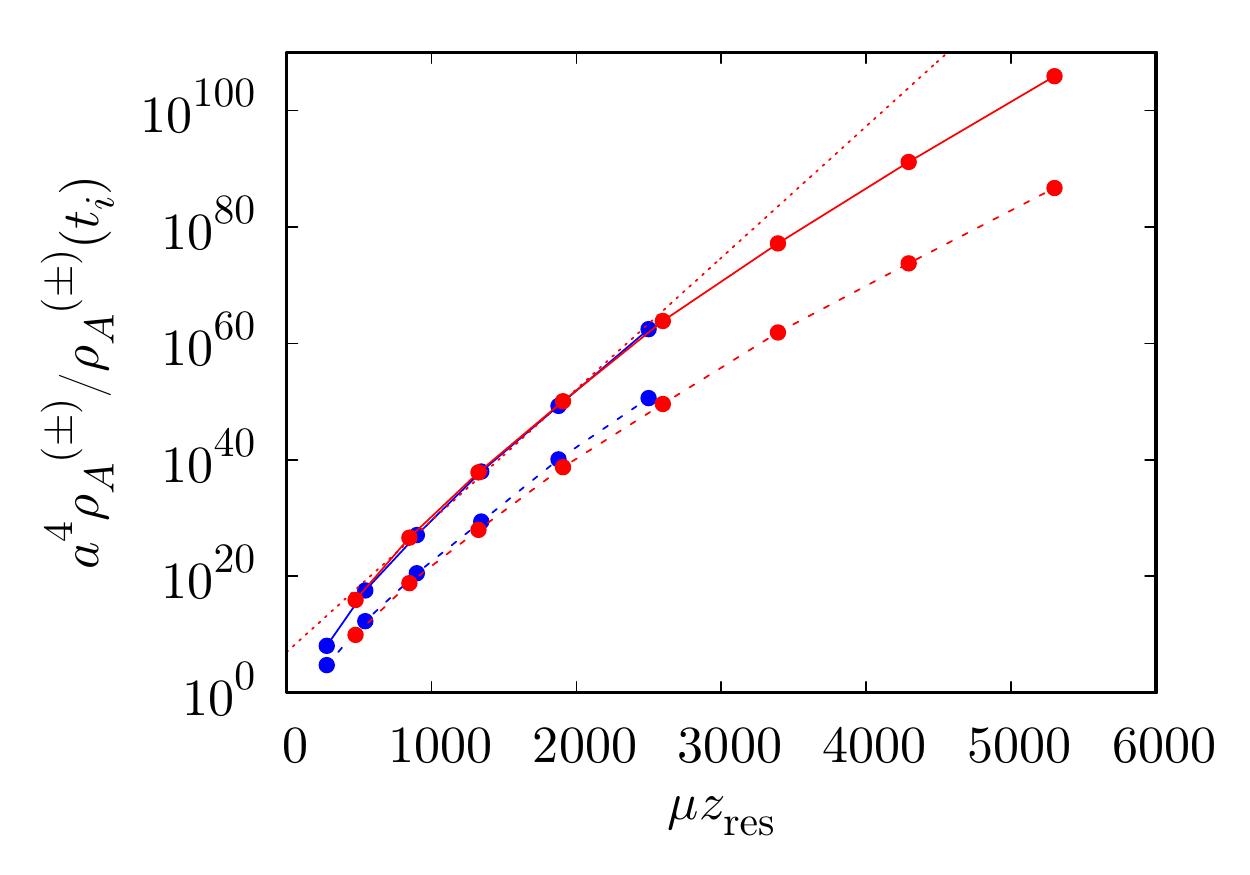}
\label{subfig:growth_notrap}
}
\subfigure[Step-function trapping, $\theta_i=1$]{
\includegraphics [width = 7.5cm, clip]{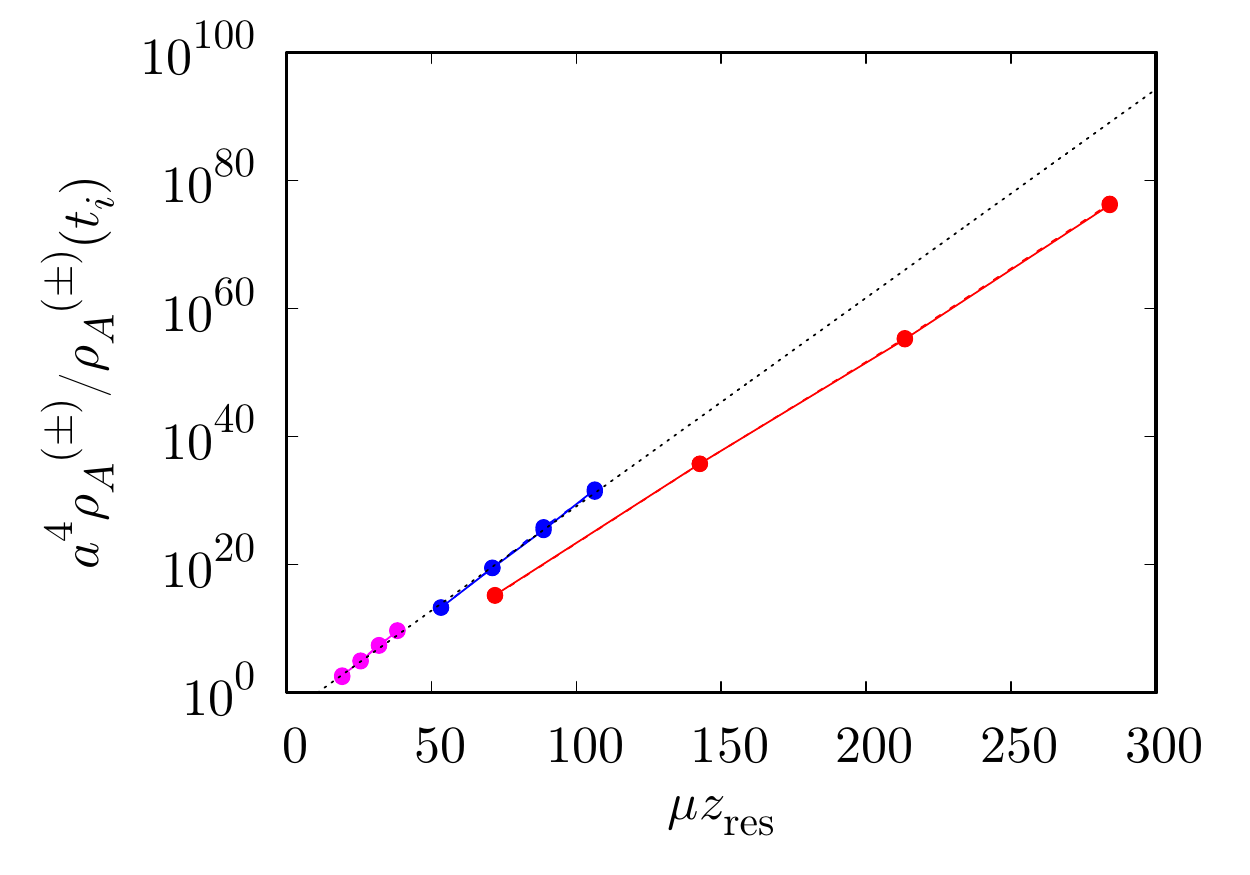}
\label{subfig:growth_a}
}
\subfigure[Trapped QCD axion, $N_H = 3$]{
\includegraphics [width = 7.5cm, clip]{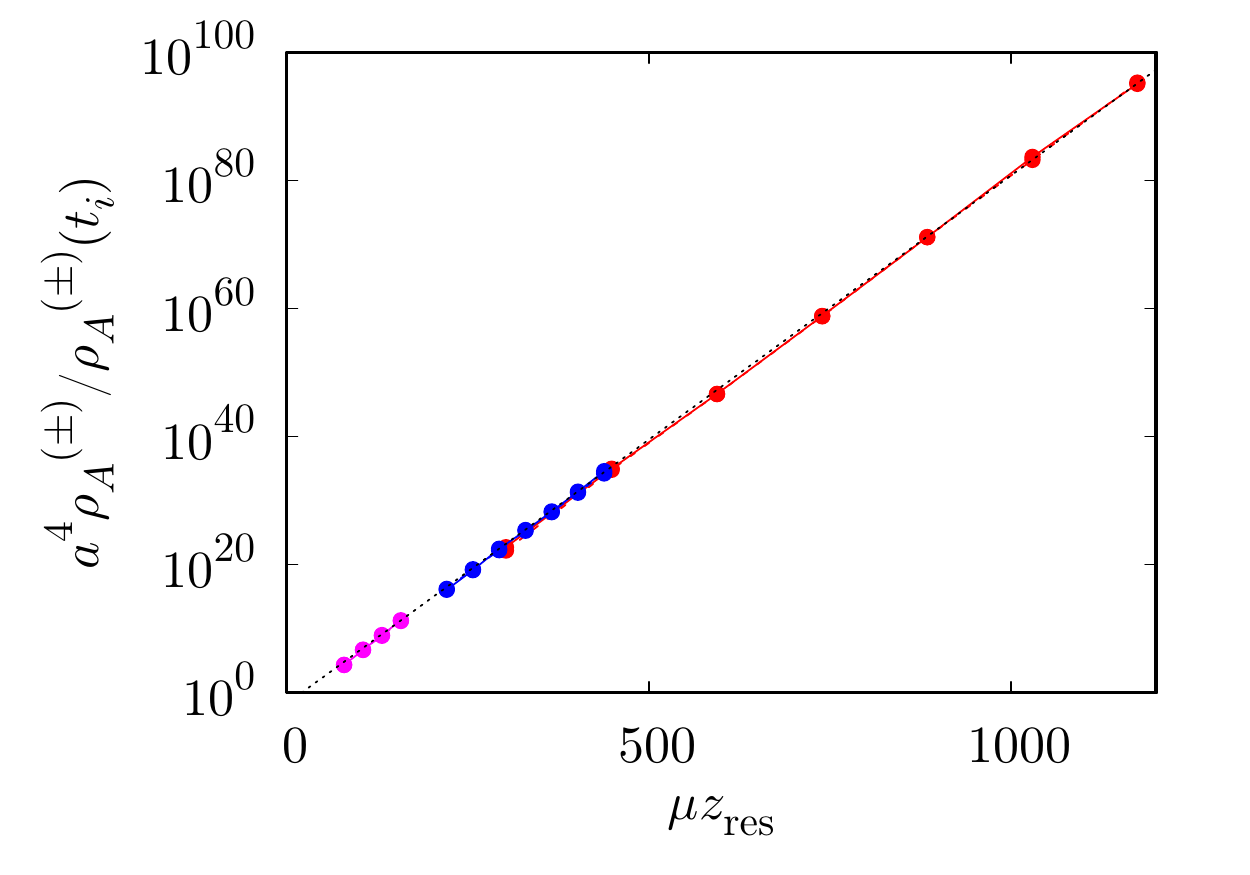}
\label{subfig:growth_b1}
}
\subfigure[Trapped QCD axion, $N_H = 6$]{
\includegraphics [width = 7.5cm, clip]{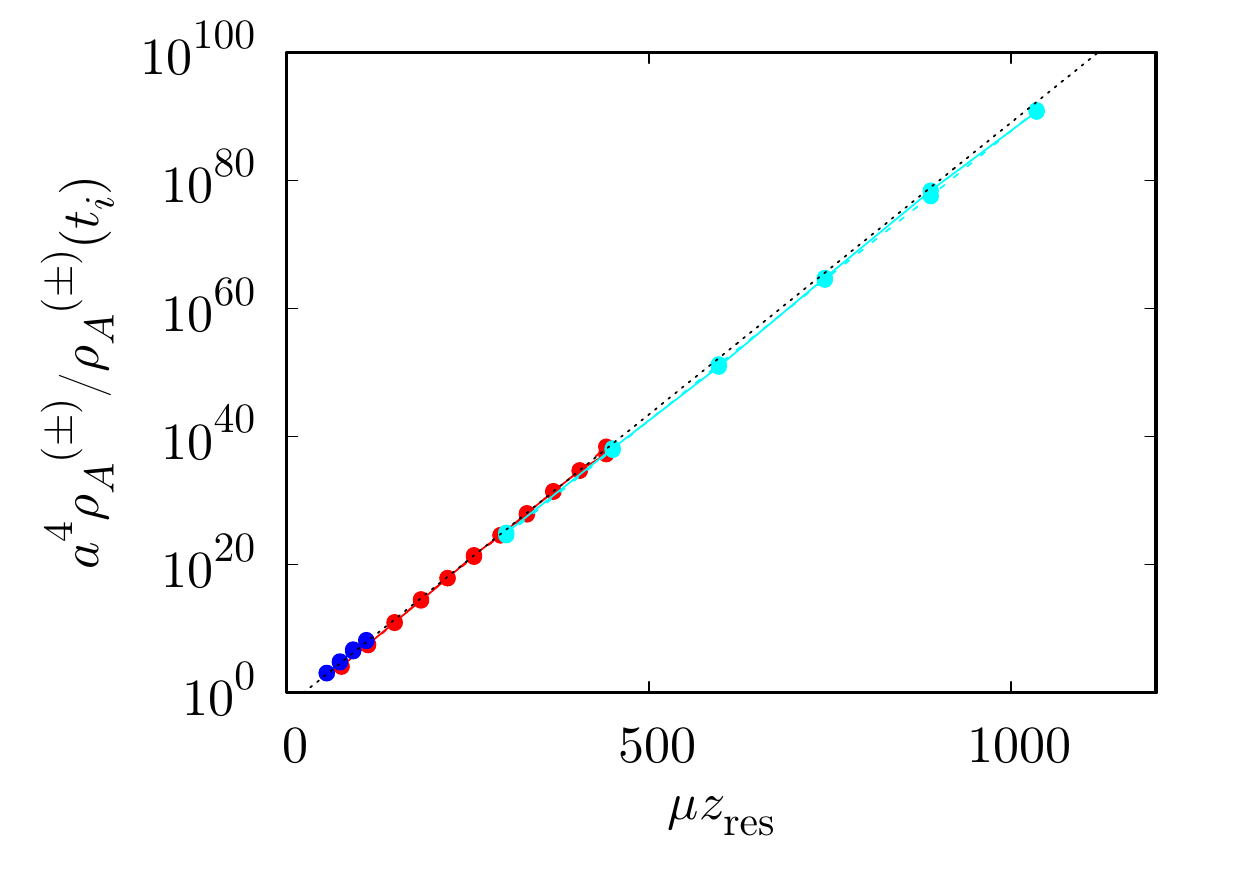}
\label{subfig:growth_b2}
}
\caption{
Total growth factor of comoving energy density of dark photons as a function of $\mu z_{\rm res}$ with $k=a_{\rm osc}m_a$ (see Eq.~(\ref{eq:muz})). 
{\it Top-left}: the case without trapping with $\theta_i = 2\pi/3$. Dotted line is a fitting function, $\exp(0.052\mu z_{\rm res})$. {\it Top-Right}: step-function trapping case with $\theta_i=1$. Dotted line is a fitting function, $\exp(0.75\mu z_{\rm res})$ {\it Bottom panels}: trapped QCD axion model with $N_H=3$ (left), 6 (right).  We have taken $\beta=1$ (red), 0.5 (blue), 0.3 (magenta) and 2 (cyan). Dotted lines are fitting formula, $\exp(0.19\mu z_{\rm res})$ (left) and $\exp(0.21\mu z_{\rm res})$ (right).
}
\label{fig:growth}
\end{figure}

Fig.~\ref{fig:growth} shows the total growth factor of the energy density of dark photons as a function of the growth exponent $\mu z_{\rm res}$ defined by Eq~(\ref{eq:muz}).
The growth can be well-fitted by the analytic formula (dotted lines) especially in the case of trapped QCD axion model.
The deviation from the fitting can be understood by the effect of the cosmic expansion.
In particular, when the Hubble parameter is comparable to the mass, the growth shows a step-like behavior due to the tachyonic instability as one can see in Fig.~\ref{subfig:ev_growth_notrap}. In this case, the growth is clearly delayed compared with the monochromatic exponential growth which we assumed in the derivation of the total growth exponent in Eq.~(\ref{eq:muz}). If the onset of the oscillation is delayed due to the trapping and the cosmic expansion becomes negligible in one oscillation period, the growth can be regarded as a monochromatic exponential function, which explains why it can be well approximated by the fitting function.

\section{Discussion and conclusions} \label{sec:disc}

We have proposed a modification to the axion dynamics in this paper to demonstrate that resonant dark photon production can occur efficiently, even for a small axion-dark photon coupling. This relies on a later onset of the axion oscillations than is typical. We have studied the time evolution of dark photons by varying the trapping duration for various coupling constants. Our results have shown that the required axion-dark photon coupling constant becomes small in proportion to the square root of the Hubble parameter at the onset of axion  oscillations, as expected from the analytical estimate of the growth rate. In other words, the total growth exponent is proportional to $\beta^2 m_a/H_{\rm osc}$ (see Eq.~(\ref{eq:muz})). This implies that the resonant production of dark photons can occur efficiently by trapping axions for an arbitrarily small coupling constant, eliminating the need for a  complicated UV completion to obtain the large axion-dark photon coupling. While the axion dynamics are more complex, only two potentials are needed, and the efficiency of dark photon production depends on when the trapping ends, which typically happens when the two potential heights become comparable.

Our mechanism is applicable to a variety of dark photon production scenarios. For example, it could be applied to scenarios where the abundance of the QCD axion with a large decay constant is suppressed by producing dark photons. In the conventional scenario where the axion starts to oscillate when the Hubble parameter is equal to the axion mass, the axion abundance is reduced to about $1/10$~\cite{Kitajima:2017peg}, but we need to run lattice numerical simulations to determine how much the axion abundance is reduced in our mechanism. In particular, because of the relative weakness of the cosmic expansion effect, the interactions between axions and dark photons may continue to be effective for a longer time after the back reaction becomes important, and may significantly change the final ratio of the axion and dark photon energy abundances. This is left for a further study.

One may apply this mechanism to the early dark energy models using an axion field~\cite{Gonzalez:2020fdy,Nakagawa:2022knn} where the axion-dark photon coupling is used either for an efficient reheating to avoid the use of contrived axion potential of the original model~\cite{Poulin:2018cxd} or for generating a heavy mass for dark Higgs. The source of the trapping potential could behave as dark radiation or hot dark matter.

Another application may be the production of the primordial magnetic field~\cite{Fujita:2015iga}. However, it relies on the inverse cascade decay of helical magnetic fields. On the other hand, in our mechanism using the trapped axion, the circular polarization asymmetry of the produced photons is significantly suppressed. Therefore, it is difficult to generate the primordial magnetic field by our mechanism.

Similarly, for gravitational waves produced by dark photons, circular polarization asymmetry is expected to be significantly reduced compared to the conventional case due to the weak asymmetry of the first few oscillations of the axion, but we need to run lattice numerical simulations to obtain a quantitative evaluation.

\section*{Acknowledgment}
F.T. thanks Tomohiro Fujita for a useful comment on the magnetogenesis. The present work is supported by the Graduate Program on  JSPS KAKENHI Grant Numbers 19H01894 (N.K.), 20H01894 (F.T.), 20H05851 (F.T. and N.K.), 21H01078 (N.K.), 21KK0050 (N.K.),
JSPS Core-to-Core Program (grant number: JPJSCCA20200002) (F.T.). This article is based upon work from COST Action COSMIC WISPers CA21106,  supported by COST (European Cooperation in Science and Technology).

\bibliographystyle{utphys}
\bibliography{ref}

\end{document}